\documentclass{aa}
\begin{document}
\newcommand\bsec{\hbox{$.\!\!{\arcsec}$}}
\newcommand\rsec{\hbox{$.\!\!{^s}$}}
\newcommand\RA[4]{#1$^{\rm h}$#2$^{\rm m}$#3\rsec#4}
\newcommand\DEC[3]{#1$^{\circ}$#2\arcmin#3\arcsec}
\newcommand{\magpt}[2]{\mbox{$\rm #1\hspace{-0.25em}\stackrel{m}{.}
   \hspace{-1.0mm}#2$}}               
\newcommand\ebv{$E_{B-V}$} 
\newcommand\teff{$T_{\rm eff}$} 
\newcommand\logt{$ \log\,T_{\rm eff}$}
\newcommand\logg{$ \log\,g$}
\newcommand\loghe{$ \log{\frac{n_{\rm He}}{n_{\rm H}}}$}
\newcommand\Msolar{${\rm M_\odot}$}
\thesaurus{01 (08.06.3, 08.16.3, 08.23.1, 10.07.3 NGC~6397)}

\title{First VLT spectra of white dwarfs in a globular cluster}
\author{S.~Moehler\inst{1}\fnmsep\thanks{Based on observations collected at the 
 European Southern Observatory (ESO~N$^{\b{o}}$~63.H-0348)} 
 \and U.~Heber\inst{1} \and R. Napiwotzki\inst{1} \and
 D. Koester\inst{2} \and A. Renzini\inst{3}}
\offprints{S.~Moehler}
\institute {Dr. Remeis-Sternwarte, Astronomisches Institut der Universit\"at
 Erlangen-N\"urnberg, Sternwartstr. 7, 96049 Bamberg, Germany 
 (e-mail: moehler,heber,napiwotzki@sternwarte.uni-erlangen.de)
\and Institut f\"ur Theoretische Physik und Astrophysik, Abteilung 
 Astrophysik, Leibnizstr. 15, D-24098 Kiel, Germany
\and European Southern Observatory, Karl Schwarzschild-Str. 2, D-85748 
Garching bei M\"unchen, Germany
}
\date{}
\maketitle


\begin{abstract}
We present the first spectra obtained with the {\em Very Large Telescope}
for white dwarfs in a
globular cluster. Estimates of atmospheric parameters are obtained and
compared to evolutionary tracks. We discuss possible implications for the
distance scale of globular clusters and white dwarf evolution and
demonstrate how white dwarfs might be used to establish an independent
distance scale to globular clusters. 
\end{abstract}


\section{Introduction} 
White dwarfs are the final stage of all low-mass stars and therefore all
single stars in a globular cluster that currently finish their
nuclear-burning lifetimes are expected to evolve into white dwarfs. As this
has been the situation for many billions of years globular clusters should
contain many white dwarfs. However, these stars managed to evade detection
until photometric white dwarf sequences in globular clusters were
discovered recently by observations with the {\em Hubble Space Telescope}
(HST) (Paresce et al.\ \cite{pade95}, Richer et al.\ \cite{rifa95},
\cite{rifa97}, Cool et al.\ \cite{copi96}, Renzini et al.\ \cite{rebr96}).
Photometric observations contain only a limited amount of information: The
two chemically distinct white dwarfs sequences (hydrogen-rich DA's and
helium-rich DB's) in principle can be distinguished by their photometric
properties alone in the temperature range $10,000\,K \leq T_{\mathrm{eff}}
\leq 15,000$\,K (see Bergeron et al.\ \cite{bewe95}). Renzini et al.\
(\cite{rebr96}) classified two white dwarfs in NGC\,6752 as DB's by this
method. However, without a spectral classification, both stars can also be
explained as high mass DA white dwarfs, possibly a product of merging.
Richer et al.\ (\cite{rifa97}) speculate that the brightest white dwarf in
M\,4 (V=22.08) might be a hot (27,000K) DB star. 
\newline\indent
The location of the white dwarf cooling sequence (and thus the brightness
of the white dwarfs) is also sensitive to the white dwarf mass. Renzini et
al.\ (\cite{rebr96}) argued that the white dwarf masses in globular
clusters are constrained to the narrow range 0.51\Msolar $\leq {\rm M_{WD}}
\leq$ 0.55\Msolar, but some systematic differences between clusters are
obvious: At a given metallicity some globular clusters (e.g.\ NGC\,6752)
possess very blue horizontal branches (HB's) with HB star masses as low as
0.50\Msolar. Such extreme HB stars evolve directly to low-mass C/O white
dwarfs (bypassing the AGB), shifting the mean white dwarf mass closer to
0.51\Msolar. Other clusters show only red HB stars, which will evolve to
the AGB and form preferably white dwarfs with masses of $\approx$0.55\Msolar. 
\begin{figure}
\vspace{8cm}
\includegraphics{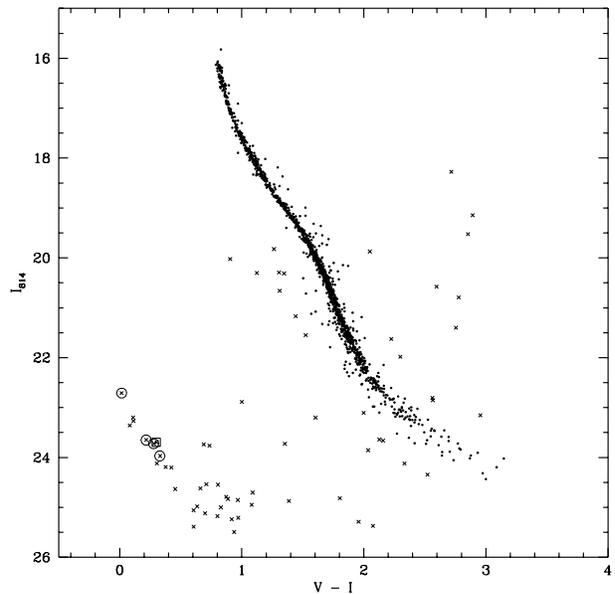}
\caption{Colour-magnitude diagram of NGC~6397 (King et al.\ \cite{kian98},
their Fig.~2). Open circles mark the four white dwarfs for which spectra 
could be obtained, the open square marks WF4-205 (see text).\label{cmd}} 
\end{figure}
Low-mass white dwarfs (M$<$0.45\Msolar) with a degenerate He core are
produced if the red giant branch evolution is terminated by binary
interaction before the helium core exceeds the minimum mass for helium
burning. Recently, Cool et al.\ (\cite{cogr98}) found 3 faint UV-bright
stars in NGC\,6397 which they suggest could be He white dwarfs (supported
by Edmonds et al.\ \cite{edgr99}). Massive white dwarfs may be produced
from blue stragglers or by collisions of white dwarf-binaries with
subsequent merging (e.g. Marsh et al.\ \cite{madh95}). 
\newline\indent
Only a detailed spectroscopic investigation can provide masses and absolute
luminosities of the individual globular cluster white dwarfs. This is also
very important for the use of white dwarfs as standard candles to derive
distances to globular clusters (Renzini et al.\ \cite{rebr96}): The basic
idea is to fit the white dwarf cooling sequence of a globular cluster to an
appropriate empirical cooling sequence of local white dwarfs with well
determined trigonometric parallaxes. The procedure is analogous to the
classical main sequence fitting but avoids the complications with
metallicity -- white dwarfs have virtually metal free atmospheres. In
addition they are locally much more abundant than metal-poor subdwarfs. The
arrival of the {\sc Hipparcos} results as well as new metallicity
determinations have rekindled the debate on globular cluster distances (see
the review by Reid \cite{reid99} and references therein). A further check
on the distance is therefore urgently needed. 
\newline\indent
We started an observing programme at the ESO {\em Very Large Telescope
(VLT)} to obtain spectra of white dwarfs in globular clusters. The
programme consists of two parts: First, low S/N ($\approx 10$) spectra of
the white dwarf candidates are obtained to verify their spectral type and
estimate their effective temperatures. In a second run we plan to observe
higher S/N ($\approx 30$) spectra that will allow to derive \logg\ with an
internal error of $\le 0.1$dex. Here we report on the very first results
for NGC~6397. 

\begin{table}
\caption{Target coordinates and photometric data (Cool priv. comm.). 
\label{targ}}
\begin{tabular}{lrrrr}
\hline
Star & $\alpha_{2000}$ & $\delta_{2000}$ & $V$ & $V-I$\\
\hline
WF4-358 & \RA{17}{40}{58}{52} & \DEC{$-$53}{42}{25.3} & \magpt{22}{73} & 
\magpt{+0}{02}\\
WF2-479 & \RA{17}{41}{07}{06} & \DEC{$-$53}{44}{45.1} & \magpt{23}{87} & 
\magpt{+0}{22}\\
WF4-205 & \RA{17}{41}{01}{58} & \DEC{$-$53}{43}{15.3} & \magpt{23}{99} &
\magpt{+0}{30}\\
WF2-51 & \RA{17}{41}{04}{52} & \DEC{$-$53}{43}{55.8} & \magpt{24}{00} & 
\magpt{+0}{28}\\
WF2-846 & \RA{17}{41}{01}{78} & \DEC{$-$53}{44}{47.2} & \magpt{24}{30} & 
\magpt{+0}{33}\\
\hline
\end{tabular}
\end{table}
\begin{figure}
\vspace{8.5cm}
\includegraphics{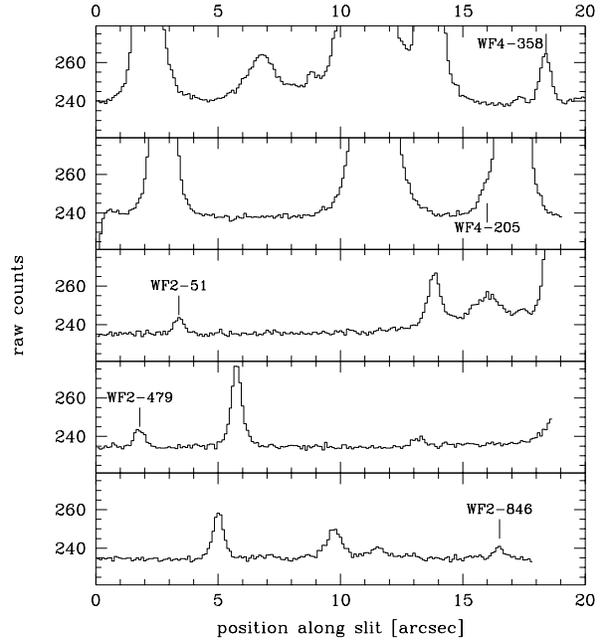}
\caption{Traces through the slits along the spatial axis. The lines mark 
the white dwarf spectra (WF4-205 unfortunately lies on the wing of a much 
brighter star).\label{slits}} 
\end{figure}

\begin{figure}
\vspace{8.5cm}
\includegraphics{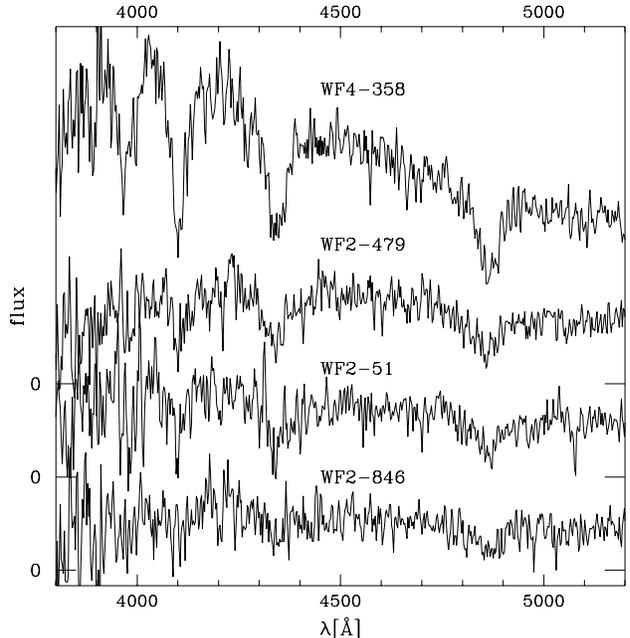}
\caption{The (relatively flux-calibrated) spectra of the white dwarfs in 
NGC~6397. The spectra of the three faintest stars 
are offset from each other as they would otherwise overlap. WF4-358 has no 
additional offset relative to WF2-479.\label{spectra}} 
\end{figure}
\section{Observations and Data Reduction}
Cool et al.\ (\cite{copi96}) discovered the white dwarfs using the {\em
Wide Field and Planetary Camera~2} (WFPC2) onboard the HST to observe the
globular cluster NGC~6397. From the improved colour-magnitude diagram of
King et al.\ (\cite{kian98}) targets brighter than $V\approx25^{\rm m}$
were selected (see Fig.~\ref{cmd}). The WFPC2 images were convolved to a
seeing of 0\bsec5 to select targets that are sufficiently uncrowded to be
observable from the ground (see Table~\ref{targ} and Fig.~\ref{slits}). The
stars were observed with the {\em FOcal Reducer/low dispersion
Spectrograph} (FORS) at Unit Telescope 1 of the VLT using the high
resolution collimator (0\bsec1/pixel) to allow better extraction of the
spectra and get a better handle on cosmic rays. We used the {\em
multi-object spectroscopy} (MOS) mode with the grism 300V and a slit width
of 0\bsec8. The slit width was chosen to be larger than the required seeing
to avoid slit losses due to imperfect pointing of the telescope. The data
were obtained in service mode under excellent conditions (seeing below
0\bsec55, no moon) with a total exposure time of 90 minutes. The final
resolution as judged from a wavelength calibration spectrum obtained with a
0\bsec5 slit is $\approx$11.5~\AA. A trace along the spatial axis of the
slitlets at about 4550~\AA\ is plotted in Fig.~\ref{slits}. Unfortunately
WF4-205 lies so close to a bright star that even at this excellent
seeing its spectrum could not be extracted. 

Due to the use of slit blades instead of fibers or masks the MOS slitlets
are very well defined and can be treated like long slits. The spectra were
therefore corrected for bias, flat-fielded, wavelength calibrated, and
extracted as described by Moehler et al.\ (\cite{mohe97}). We find only a
diffuse and rather low sky background without any strong sky lines below
5150~\AA. The spectra were relatively flux calibrated using LTT~7987
(Hamuy et al.\ \cite{hawa92}) and are plotted in Fig.~\ref{spectra}. All
four stars display only strong broad Balmer lines, which is characteristic
for hydrogen-rich white dwarfs (DA stars). 

\section{Atmospheric parameters}
Although the white dwarf spectra have low signal-to-noise, they are
sufficient for rough parameter estimates. The atmospheric parameters are
obtained by simultaneously fitting profiles of the observed Balmer lines
with model spectra using the least-square algorithm of Bergeron et al.\
(\cite{besa92}; see Napiwotzki et al.\ 1999 for minor modifications).
Analyses were performed with Koester's LTE models as described in Finley et
al.\ (\cite{fiko97}). As a check we repeated the analysis of the hottest
star in our sample (WF4-358) with the non-LTE grid described in Napiwotzki
et al.\ (\cite{nagr99}). Since the non-LTE code does not treat convection
and ignores molecular opacities reliable atmospheric models cannot be
calculated for the three cooler white dwarfs. 

\begin{figure}
\vspace{8.cm}
\includegraphics{b1293.f4}
\caption{Sample fits to the spectra of WF4-358 (\teff\ =
18,200~K, \logg\ = 7.3) and WF2-479 (\teff\ = 11,000~K, \logg\= = 7.7).
\label{wd_fit}} 
\end{figure}

\begin{table}[h]
\caption{Effective temperatures and $\chi^2$ values for the white dwarfs
from the fit of H$_\delta$, H$_\gamma$, H$_\beta$ at fixed \logg\ (see text
for details). Also given are masses derived from theoretical tracks of
Bl\"ocker (\cite{bloe95}) and Driebe et al.\ (\cite{drsch98}) and absolute
magnitudes (Bergeron et al.\ \cite{bewe95}). Using Bergeron et al.'s
(\cite{bewe95}) Table~3 we derived $T_{\rm eff, (V-I)_0}$ from $(V-I)_0$,
which was calculated from $V-I$ assuming $E_{V-I}$ = \magpt{0}{225}.
\label{tab-results}} 
\begin{tabular}{lrrrrr}
\hline
star & $\chi^2$ & \teff & $T_{\rm eff, (V-I)_0}$ & M & $M_V$ \\
   & & [K] & [K] & [\Msolar] & \\
\hline
\multicolumn{6}{c}{\logg\ = 7.5}\\
\hline
WF4-358 & 1.04 & 18900 & & 0.42 & \magpt{10}{0} \\
WF2-479 & 1.15 & 10800 & & 0.39 & \magpt{11}{1} \\
WF2-51 & 1.46 & 10900 & & 0.39 & \magpt{11}{1} \\
WF2-846 & 0.71 & 10500 & & 0.38 & \magpt{11}{2} \\
\hline
\multicolumn{6}{c}{\logg\ = 7.7}\\
\hline
WF4-358 & 1.07 & 19500 & & 0.48 & \magpt{10}{3} \\
WF2-479 & 1.14 & 11000 & & 0.46 & \magpt{11}{4} \\
WF2-51 & 1.46 & 11100 & & 0.46 & \magpt{11}{4} \\
WF2-846 & 0.71 & 10600 & & 0.46 & \magpt{11}{5} \\
\hline
\multicolumn{6}{c}{\logg\ = 8.0}\\
\hline
WF4-358 & 1.13 & 20300 & 19400 & 0.62 & \magpt{10}{7} \\
WF2-479 & 1.14 & 11200 & 11500 & 0.59 & \magpt{11}{8} \\
WF2-51 & 1.47 & 11300 & 10800 & 0.59 & \magpt{11}{8} \\
WF2-846 & 0.71 & 10800 & 10300 & 0.59 & \magpt{11}{9} \\
\hline
\end{tabular}\\
\end{table}
Fitting the lines H$_\beta$ to H$_\epsilon$ for WF4-358 (see
Fig.~\ref{wd_fit}) gives 18,200$\pm$1300~K and 7.30$\pm$0.36 for \teff\ and
\logg, respectively ($\chi^2$ = 0.93). The errors given here are 1$\sigma$
errors obtained from the $\chi^2$ fit. Omitting H$_\epsilon$ from the fit
results in 17,800~K and 7.19 ($\chi^2$ = 1.02) with more or less unchanged
errors. The results of the non-LTE analysis are essentially identical to
those obtained with Koester's models, differing only by small fractions of
the formal errors ($\Delta$\teff $\approx$500\,K, $\Delta$\logg $\approx$
0.07\,dex). The surface gravity is surprisingly low and suggests that
WF4-358 could be a bright ($M_V$=\magpt{9}{7}) helium white dwarf of
(0.36$\pm$0.12)\Msolar. Within the error bars, however, the derived
parameters are also consistent with a low-mass C/O white dwarf. For the
remaining three stars the S/N is too low to determine \teff\ and \logg\
simultaneously. We thus fitted H$_\delta$, H$_\gamma$, H$_\beta$
(H$_\epsilon$ being too noisy) for WF2-51, WF2-479, and WF2-846 for three
fixed values of \logg\ (8.0, 7.7, 7.5, see Fig.~\ref{wd_fit} for an
example). These \logg\ values correspond to C/O white dwarfs of $\approx$
0.6\Msolar, low-mass C/O white dwarfs of $\approx$0.5\Msolar, and He white
dwarfs of $\approx$0.4\Msolar, respectively (see below). The formal errors
are $\approx$550~K (WF2-479), $\approx$650~K (WF2-846, WF4-358), and
$\approx$790~K (WF2-51). The errors for the cooler stars are relatively
small despite their low S/N because -- at fixed \logg\ -- the line profiles
are much more sensitive to temperature variations at \teff
$\approx$11,000~K than at \teff $\approx$18,000~K. The relatively large
$\chi^2$ value for WF2-51 suggests that either the noise in this spectrum
has been underestimated or that the spectrum contains additional features
that are not well described by the model spectra. The temperatures derived
from the Balmer lines agree quite well with those obtained from $(V-I)_0$
using the theoretical colours of Bergeron et al.\ (\cite{bewe95}, \logg\ = 
8.0). 

The masses given in Table~\ref{tab-results} were derived by interpolation
between the evolutionary tracks of C/O white dwarfs calculated by Bl\"ocker
(\cite{bloe95}) and the He white dwarf tracks of Driebe et al.\
(\cite{drsch98}). Finally, absolute magnitudes $M_V$ were calculated for
each parameter set with the photometric calibration of Bergeron et al.\
(\cite{bewe95}). 

\section{The distance to NGC\,6397} 
The old distance modulus to NGC\,6397 was $(m-M)_0$ = \magpt{11}{71} with a
reddening of \ebv\ of \magpt{0}{18} (Djorgovski \cite{djor93}). Using local
metal-poor subdwarfs to fit the main sequence of NGC\,6397 Reid \& Gizis
(\cite{regi98}) obtained a mean distance modulus
of $(m-M)_0$ = \magpt{12}{20}$\pm$\magpt{0}{15} for
\ebv\ = \magpt{0}{19}. Thus NGC\,6397 is a good example for the large
differences between old and new distances to globular clusters. The
absolute magnitudes given in Table~\ref{tab-results} for \logg\ = 7.5, 7.7,
8.0 (M$_{\rm WD}$ = 0.4\Msolar, 0.5\Msolar, 0.6\Msolar) yield mean true
distance moduli (for \ebv\ = \magpt{0}{18}) $(m-M)_0$ of \magpt{12}{3},
\magpt{12}{0}, and \magpt{11}{6}, respectively, with an r.m.s. error of
\magpt{0}{17}. Considering the error bars of the various distance
determinations the long distance scale would be more consistent with white
dwarf masses $\le$0.5\Msolar\ and the short distance scale with masses
$>$0.5\Msolar. The longer distance moduli obtained for low-mass white
dwarfs also result in masses for blue HB stars (Heber et al.\
\cite{hemo97}) and a hot post-AGB star (ROB\,162, Heber \& Kudritzki
\cite{heku86}) that agree with canonical evolutionary theory. 

The distance moduli derived for a given \logg\ from Tables~\ref{targ} and
\ref{tab-results} show a systematic variation with the brightest star
(WF4-358) yielding the smallest distance and the faintest star (WF2-846)
giving the largest distance. This variation could reflect the fact that the
stars may not all have the same surface gravity: From their different
apparent magnitudes (i.e. different absolute magnitudes) it is plausible
that WF4-358 has the smallest \logg\ and WF2-846 the largest. The quality
of the current data, however, does not allow to verify this idea. 

\section{Conclusions} 
Using VLT-FORS1 multi object spectroscopy we have confirmed four white
dwarf candidates to be hydrogen-rich DA white dwarfs. The gravity
determined for the brightest star, WF4-358, suggests that it could be a He
white dwarf with a mass of (0.36$\pm$0.12)\Msolar, although the error bars
are large enough to also accommodate a C/O white dwarf. Temperatures derived
for the three cooler and fainter stars for fixed \logg\ would put them near
the red edge of the ZZ Ceti instability strip, for which Bergeron et al.\
(\cite{bewe95b}) determine a temperature range of 11,160~K to 12,460~K
using their preferred ML2/$\alpha$=0.6 prescription for the treatment of
convection. Therefore, a search for photometric variability of WF2-479 and
WF2-51 -- if successful -- could place important additional constraints on
these stars. The systematic variation of the distance moduli derived for a
given \logg\ shows that the assumption of a constant mass for
all white dwarfs in a globular cluster may bias a distance determination.
However, due to the low S/N of the current data, these results are
preliminary. Once higher quality spectra are available, which will allow
more accurate parameter (and thus mass) determinations, analyses of white
dwarfs in globular clusters will become a powerful tool for independent
distance estimates. 

\acknowledgements
We highly appreciate the work performed by the FORS team in building an
excellent instrument and the efforts of the staff at the ESO Paranal
observatory and ESO Garching that made these observations possible. We
thank Dr.\ A.\ Cool for providing us with the photometry and coordinates of
the white dwarf candidates and an anonymous referee for valuable comments.
S.M. acknowledges financial support from the DARA under grant
50~OR~96029-ZA. 

{}

\begin{thebibliography}{}
\bibitem[1992]{besa92}
Bergeron P., Saffer R.A., Liebert J., 1992, ApJ 394, 228
\bibitem[1995a]{bewe95}
Bergeron P., Wesemael F., Beauchamp A., 1995a, PASP 107, 1047
\bibitem[1995b]{bewe95b}
Bergeron P., Wesemael F., Lamontagne R., et al., 1995b, ApJ 449, 258
\bibitem[1995]{bloe95}
Bl\"ocker T., 1995, A\&A 299, 755
\bibitem[1996]{copi96}
Cool A.M., Piotto G., King I.R., 1996, ApJ 468, 655
\bibitem[1998]{cogr98}
Cool A.M., Grindlay J.E., Cohn H.N., Lugger P.M., Bailyn C.D., 1998, ApJ 
508, L75
\bibitem[1993]{djor93}
Djorgovski S., 1993, in {\it Structure and Dynamics of Globular Clusters},
  eds. S.G. Djorgovski \& G. Meylan, ASP Conf. Ser. 50 (San Francisco),
  p. 373
\bibitem[1998]{drsch98}
Driebe T., Sch\"onberner D., Bl\"ocker T., Herwig F., 1998, A\&A 339, 123
\bibitem[1999]{edgr99}
Edmonds P.D., Grindlay J.E., Cool A., et al., 1999, ApJ 516, 250
\bibitem[1997]{fiko97}
Finley D.S., Koester D., Basri G., 1997, ApJ 488, 375
\bibitem[1992]{hawa92}
Hamuy M., Walker A.R., Suntzeff N.B., et al., 1992, PASP 104, 533
\bibitem[1986]{heku86}
Heber U., Kudritzki R.P., 1986, A\&A 169, 244
\bibitem[1997]{hemo97}
Heber U., Moehler S., Reid I.N., 1997, ESA SP-402, p.461 
\bibitem[1998]{kian98}
King I.R., Anderson J., Cool A.M., Piotto G., 1998, ApJ 492, L37
\bibitem[1995]{madh95}
Marsh T.R., Dhillon V.S., Duck S.R., 1995, MNRAS 275, 828
\bibitem[1997]{mohe97}
Moehler S., Heber U., Rupprecht G., 1997, A\&A 319, 109
\bibitem[1999]{nagr99}
Napiwotzki R., Green P.J., Saffer R.A., 1999, ApJ 517,399
\bibitem[1995]{pade95}
Paresce F., de Marchi G., Romaniello M., 1995, ApJ 440, 216
\bibitem[1999]{reid99}
Reid I.N., 1999, ARAA 37, 191
\bibitem[1998]{regi98}
Reid I.N., Gizis J.E., 1998, AJ 116, 2929
\bibitem[1996]{rebr96}
Renzini A., Bragaglia A., Ferraro F.R., et al., 1996, ApJ 465, L23
\bibitem[1995]{rifa95}
Richer H.B., Fahlmann G.G., Ibata R.A., et al., 1995, ApJ 451, L17
\bibitem[1997]{rifa97}
Richer H.B., Fahlmann G.G., Ibata R.A., et al., 1997, ApJ 484, 741
\end{thebibliography}
\end{document}